\documentclass[a4paper,12pt,axodraw]{revtex4}
\usepackage{axodraw}
\usepackage{slashed}
\usepackage{latexsym}
\usepackage{amssymb}
\usepackage{amsmath,amssymb,amsthm,graphicx,bm,color}

\renewcommand{\Delta}{\varDelta} 
\renewcommand{\Gamma}{\varGamma} 
\renewcommand{\Omega}{\varOmega} 
\renewcommand{\Phi}{\varPhi} 
\renewcommand{\Psi}{\varPsi} 
\renewcommand{\Sigma}{\varSigma} 
\renewcommand{\Theta}{\varTheta} 
\renewcommand{\epsilon}{\varepsilon}

\begin{document}

\title{Proton Decay Constraints on Low Scale AdS/CFT Unification}

\author{James B. Dent}

\email{james.b.dent@vanderbilt.edu}

\author{Thomas W. Kephart}

\email{tom.kephart@gmail.com}

\affiliation{Department of Physics and Astronomy, Vanderbilt
University, Nashville, TN 37235} 

\date{\today}

\begin{abstract}
Dark matter candidates and proton decay in a class of  models based on the AdS/CFT correspondence are discussed. We show that the present bound on the proton decay lifetime is inconsistent with ${\cal N}  = 1$ SUSY, and strongly constrains ${\cal N}  = 0$ non-SUSY, low scale trinification type unification of orbifolded AdS$\otimes S^5$ models.
\end{abstract}


\maketitle

\newpage

\section{Introduction}

Models with low scale ($\sim$ TeV) unification are of great potential interest for LHC physics since the unification scale, and thus any new particles that exist at that scale, will be well within the reach of the LHC.  Many models based on the AdS/CFT correspondence are of this type. When orbifolded they can lead to products of $SU(N)$ gauge groups\cite{Lawrence:1998ja}, 
(Actually they are products of $U(N)$ gauge groups, and the extra $U(1)$s can be anomalous.
  However, counter terms can be added the Lagrangian to cancel these anomalies. See  \cite{DiNapoli:2006wz} and references therein. In what follows, we will suppress the $U(1)$ factors, and assume they decouple from the analysis.)
with particle representations that can be identified with the Standard Model (SM) when the gauge group $G$ is broken to $SU(3)\times SU(2)\times U(1)$. The simplest examples are $AdS\otimes S^5/Z_n$ models with $N=3$ so $G= SU(3)^n$  where the matter fields are all in bifundamental representations. ( For $n=3$ this leads to a three family ${\cal N}=1$ trinification model \cite{Kachru:1998ys}. For a general classification of both ${\cal N}=1$ and ${\cal N}=0$ $Z_n$ orbifolded models see \cite{Kephart:2001qu,Kephart:2004qp}.)  While some progress has been made \cite{Frampton:2003cp, Frampton:2003jx, Frampton:2004xb}, the full phenomenological implications of these models are not known as detailed model building has been lacking.  Recently, however, some attempts have been made to accommodate a dark matter candidate \cite{Frampton:2006va}, the lightest conformal particle (LCP), and a stable proton  \cite{Frampton:2006pi} by the (well motivated yet ad hoc) assignment of discrete charges to particles, which thereby forbids any harmful interactions.  Here we will investigate whether such assignments can arise naturally via discrete subgroups upon spontaneous symmetry breaking, and determine their consequences.  

We will carry out our analysis within a single representative model, and then show how the results generalize. We begin with a review of the simplest non-SUSY low scale example based on a $Z_7$ orbifold \cite{Frampton:1999nb}. We then follow the quantum numbers through the symmetry breaking from $G = SU(3)^7$ to the trinification \cite{willenbrock:2006} group $SU_C(3)\times SU_L(3)\times SU_R(3)$, then to the group that contains $B-L$ symmetry, $SU_C(3)\times SU_L(2)\times U_Y(1)\times U_{B-L}(1)\times U_X(1)$, and  finally to the standard model. We show that a dark matter candidate LCP can only arise if an appropriate R-parity assignments  \cite{Frampton:2006va} is  realized phenomenologically by adding extra scalars in this model.   We also  show that dimension six proton decay operators are present and therefore proton decay occurs much too rapidly in this model. This is always the case for any low scale ${\cal N}  = 1$ SUSY trinification model of this type. There is a possible escape in the non-SUSY case, but it will take clever model building to achieve an acceptable result. The set of charge assignments that avoids proton decay given in \cite{Frampton:2006pi} is such a case, but it is not realized in the $Z_7$ orbifold model studied here.\\

\section{ non-SUSY $Z_7$}
We begin with a short review of the non-SUSY $Z_7$ orbifold model \cite{Frampton:1999nb}, but before doing so we will first pause to recall how models are constructed by orbifolding $AdS\otimes S^5$  \cite{Lawrence:1998ja}. When we desire an ${\cal N}=1$ supersymmetric theory, we embed the orbifolding group $\Gamma$ in the $\bm 4$ of the $SU(4)$ R-symmetry of the underlying ${\cal N}=4$ theory via ${\bm 4}=(a_1,a_2,a_3,a_4)$, where $a_1=1$ and the other $a_j$ are nontrivial elements of $\Gamma$. We will only be concerned with abelian $Z_n$ choices for $\Gamma$ which generate chiral supermultiplet matter fields in the bifundamental representations of the gauge group $SU^n(N)=SU_1(N)\times SU_2(N)\times ....SU_n(N) $. When $\Gamma$ is $Z_n$ the $a_j$ are of the form $a_j=e^{\frac{2\pi i n_j}{n}}$ and the matter bifundamentals are $(N,\bar{N})$ representations of $SU_i(N)\times SU_{i+n_j}(N)$ for all $i=1,2,...n$ and $n_j$. (Note, $n_1=0$ in the ${\cal N}=1$ SUSY case.) If we wish to break all the supersymmetry, $a_1$ must also be nontrivial, and therefore $n_1\neq 0$. The fermions are still generated by the embedding of $Z_n$ in the
$\bm 4$ of the initial $SU(4)$ R-symmetry, but now the scalars arise from the embedding of  $Z_n$ in the
$\bm 6=(\bm 4\times \bm 4)_A$, generated from the embedding of the $\bm 4$. The scalar bifundementals are in $(N,\bar{N})$ representations of $SU_i(N)\times SU_{i+p_k}(N)$, where the $p_k$ are obtained from the six antisymmetric combinations $e^{\frac{2\pi i p_k}{n}} =e^{\frac{2\pi i (n_j+n_j')}{n}}$. We now have sufficient background to write down the non-SUSY $Z_7$ model on which we will focus most of our attention.

\subsection{ Initial Particle Content and $\sin^2\theta_W$}

The fermions are given by the embedding of the $Z_7$ orbifolding group in the  \textbf{4} of the $SU(4)$ $R$ symmetry: ${\bf 4}=(\alpha,\alpha,\alpha^2,\alpha^3)$ \cite{Frampton:1999nb} where $\alpha=e^{\frac{2\pi i}{7}}$, and the scalars can be found from ${\bf 6}=(  \alpha^2,\alpha^3,\alpha^3,\alpha^4,\alpha^4,\alpha^5)$,
  which leads to the particle content

\begin{center}
\begin{tabular}{|l|l|l|}
\hline
\multicolumn{3}{|c|}{\textbf{Fermions}}\\
\hline
2(3,$\bar{3}$,1,1,1,1,1)  &  (3,1,$\bar{3}$,1,1,1,1)  &  (3,1,1,$\bar{3}$,1,1,1)\\
2(1,3,$\bar{3}$,1,1,1,1)  &  (1,3,1,$\bar{3}$,1,1,1)  &  (1,3,1,1,$\bar{3}$,1,1)\\
2(1,1,3,$\bar{3}$,1,1,1)  &  (1,1,3,1,$\bar{3}$,1,1)  &  (1,1,3,1,1,$\bar{3}$,1)\\
2(1,1,1,3,$\bar{3}$,1,1)  &  (1,1,1,3,1,$\bar{3}$,1)  &  (1,1,1,3,1,1,$\bar{3}$)\\
2(1,1,1,1,3,$\bar{3}$,3)  &  (1,1,1,1,3,1,$\bar{3}$)  &  ($\bar{3}$,1,1,1,3,1,1)\\
2(1,1,1,1,1,3,$\bar{3}$)  &  ($\bar{3}$,1,1,1,1,3,1)  &  (1,$\bar{3}$,1,1,1,3,1)\\
2($\bar{3}$,1,1,1,1,1,3)  &  (1,$\bar{3}$,1,1,1,1,3)  &  (1,1,$\bar{3}$,1,1,1,3)\\
\hline
\end{tabular}
\end{center}

\begin{center}
\begin{tabular}{|l|l|l|l|}
\hline
\multicolumn{4}{|c|}{\textbf{Scalars}}\\
\hline
(3,1,$\bar{3}$,1,1,1,1)  &  2(3,1,1,$\bar{3}$,1,1,1)  &  2(3,1,1,1,$\bar{3}$,1,1)  &  (3,1,1,1,1,$\bar{3}$,1)\\
(1,3,1,$\bar{3}$ 1,1,1)  &  2(1,3,1,1,$\bar{3}$,1,1)  &  2(1,3,1,1,1,$\bar{3}$,1)  &  (1,3,1,1,1,1,$\bar{3}$)\\
(1,1,3,1,$\bar{3}$ 1,1)  &  2(1,1,3,1,1,$\bar{3}$,1)  &  2(1,1,3,1,1,1,$\bar{3}$)  &  ($\bar{3}$,1,3,1,1,1,1)\\
(1,1,1,3,1,$\bar{3}$,1)  &  2(1,1,1,3,1,1,$\bar{3}$)  &  2($\bar{3}$,1,1,3,1,1,1)  &  (1,$\bar{3}$,1,3,1,1,1)\\
(1,1,1,1,3,1,$\bar{3}$)  &  2($\bar{3}$,1,1,1,3,1,1)  &  2(1,$\bar{3}$,1,1,3,1,1)  &  (1,1,$\bar{3}$,1,3,1,1)\\
($\bar{3}$,1,1,1,1,3,1)  &  2(1,$\bar{3}$,1,1,1,3,1)  &  2(1,1,$\bar{3}$,1,1,3,1)  &  (1,1,1,$\bar{3}$,1,3,1)\\
(1,$\bar{3}$,1,1,1,1,3)  &  2(1,1,$\bar{3}$,1,1,1,3)  &  2(1,1,1,$\bar{3}$,1,1,3)  &  (1,1,1,1,$\bar{3}$,1,3)\\
\hline
\end{tabular}
\end{center}

At the unification scale
$$\sin^2(\theta_W)=\frac{2}{7}$$ where $SU_L(2)\subset SU^p(3)$, $U_Y(1)\subset SU^p(3)$, and $p$ and $q$ are from the seven initial  $SU(3)s$ \cite{Kephart:2001qu}. 
We find this by starting with the gauge group $SU^n(3)$ and breaking to $SU_C(3)\times SU_L(3)\times SU_R(3)$ where $SU_C(3)\subset SU^r(3)$, $SU_L(3)\subset SU^p(3)$, and $SU_R(3)\subset SU^q(3)$ where
$n=p+q+r$, then
writing $Y$ as the sum of  diag($\frac{1}{2}$,$\frac{1}{2}$,$-1$)
  of $SU_L(3)$ and diag($1$,$1$,-2) of $SU(3)_R$ we get 
  $$sin^2\theta_W=\frac{3}{3+5\left(  \alpha_2/\alpha_Y\right)}
  =\frac{3}{3+5\left(\frac{3p}{p+2q}\right)}$$
at the unification scale. For the model at hand, $p=2$ and $q=1$, hence the result 2/7.

\subsection{ Symmetry Breaking}

We proceed as in \cite{Frampton:1999wz} by giving a VEV successively to (1,3,1,$\bar{3}$,1,1,1), (1,1,3,$\bar{3}$,1,1), (1,1,3,$\bar{3}$,1), and (1,1,3,$\bar{3}$).  Each VEV breaks an $SU(3)\times SU(3)$ pair to its diagonal $SU(3)$ subgroup and thus, after all four of these scalars have obtained VEVs, the gauge group is $SU_C(3)\times SU_L(3)\times SU_R(3)$ with three massless chiral families of fermions in complex representations 
$$3[(3,\bar{3},1)+(1,3,\bar{3})
+3(\bar{3},1,3)]_F$$
plus fermions in real representations that acquire heavy (unification scale) masses,
as well as various scalars given by
$$[19(1,1,1)  +
(1,8,1)  +
9(1,1,8) 
+3(3,\bar{3},1)  + 2(\bar{3},3,1) 
+8[(1,3,\bar{3})  + (1,\bar{3},3)]  
+4(\bar{3},1,3)  + 3(3,1,\bar{3})]_S  $$

We can then break $SU_L(3)$ to $SU_L(2)\times U_{L8}(1)$ (the subscript ``8'' refers to the $\lambda_8$ generator of $SU_L(3)$) by giving a VEV to the neutral singlet in the adjoint
%$2_3$ and $2_{-3}$ which come from the decomposition of the 8 = $1_0 + 2_{3} + 2_{-3} + 3_0
(1,8,1) which decomposes as $8 = 1_0 + 2_{3} + 2_{-3} + 3_0$ under $SU_L(2)\times U_{L8}(1)$ (the 2s are the fundamental $SU_L(2)$ representation and the subscript refers to the $U_{L8}(1)$ charge).  We can also break $SU_R(3)$ to $U_{R3}(1)\times U_{R8}(1)$ by first breaking to  $SU_R(2)\times U_{R8}(1)$ as just described (except that the 8 is now chosen from one of the nine (1,1,8) multiplets) and then giving a VEV to a neutral $U_{R3}(1)\times U_{R8}(1)$ singlet (which arises from the decomposition of an  $SU_R(2)$ adjoint, which is $3_0 = 1_2 +1_{-2} + 1_0$ under $U_{R3}(1)\times U_{R8}(1)$).  

The quantum numbers for hypercharge ($Y$), baryon-minus-lepton ($B-L$), and $X$ can be written in terms of the $U(1)$ charges $T_{L,R}^{3,8}$ (i.e., the Cartan subalgebra charges) as follows
\begin{eqnarray}
Y &=& -\frac{1}{6}T_{L}^8 + \frac{1}{2}T_{R}^8 -\frac{1}{6}T_{R}^3\\
B - L &=& -\frac{1}{3}T_{L}^8 -\frac{1}{3}T_{R}^3\\
X &=& -T_{L}^8 + 2T_{R}^3
\end{eqnarray} 

\subsection{ Low Energy Particle Content }
The gauge group is now $SU_C(3)\times SU_L(2)\times U_Y(1)\times U_{B-L}(1)\times U_X(1)$ and the particle content is given by (listing only light fermions, but all the scalars):

\begin{center}
\begin{tabular}{|l|l|l|}
\hline
\multicolumn{3}{|c|}{\textbf{Family Fermions}}\\
\hline
\hline
$Q$: $3(3,2)_{\frac{1}{6}, \frac{1}{3}, 1}$ & L: $3(1,2)_{-\frac{1}{2},-1,3}$ & $\bar{u}$: $3(\bar{3},1)_{-\frac{2}{3},-\frac{1}{3},2}$\\
$\bar{d}$:  $3(\bar{3},1)_{\frac{1}{3},-\frac{1}{3},2}$ & $\bar{e}$: $3(1,1)_{1,1,0}$ & $N$: $3(1,1)_{0,1,0}$ \\
\hline
\multicolumn{3}{|c|}{\textbf{Exotic Fermions}}\\
\hline
\hline
$\bar{h}$:  $3(\bar{3},1)_{\frac{1}{3},\frac{2}{3},-4}$& $h$:  $3(3,1)_{-\frac{1}{3},-\frac{2}{3},-2}$ & \\
$\bar{E}$:  $3(1,2)_{\frac{1}{2},0,-3}$ & $E$:  $3(1,2)_{-\frac{1}{2},0,-3}$ & $S$: $3(1,1)_{0,0,6}$\\
\hline
\end{tabular}
\end{center}

\begin{center}
\begin{tabular}{|c|c|c|c|}
\hline
\multicolumn{4}{|c|}{\textbf{Scalars}}\\
\hline
\hline 
8(1,1)$_{1,0,0}$ & 8(1,1)$_{0,1,0}$ & 8(1,1)$_{0,1,-6}$ & 8(1,1)$_{1,1,-6}$
\\ 
8(1,1)$_{-1,0,0}$ & 8(1,1)$_{0,-1,0}$ & 8(1,1)$_{0,-1,6}$ & 8(1,1)$_{-1,-1,6}
$ \\ 
8(1,1)$_{0,0,6}$ & 8(1,1)$_{1,1,0}$ & 8(1,2)$_{\frac{1}{2},0,3}$ & 8(1,2)$_{%
\frac{1}{2},0,-3}$ \\ 
8(1,1)$_{0,0,-6}$ & 8(1,1)$_{-1,-1,0}$ & 8(1,2)$_{-\frac{1}{2},0,-3}$ & 
8(1,2)$_{-\frac{1}{2},0,3}$ \\ 
8(1,2)$_{\frac{1}{2},1,-3}$ & 3(3,1)$_{-\frac{1}{3},-\frac{2}{3},-2}$ & 
3(3,1)$_{\frac{2}{3},\frac{1}{3},-2}$ & 3(3,1)$_{-\frac{1}{3},-\frac{2}{3},4}
$ \\ 
8(1,2)$_{-\frac{1}{2},-1,3}$ & 2(\={3},1)$_{-\frac{1}{3},-\frac{2}{3},-2}$ & 
4(\={3},1)$_{-\frac{2}{3},-\frac{1}{3},2}$ & 4(\={3},1)$_{\frac{1}{3},\frac{2%
}{3},-4}$ \\ 
3(3,1)$_{-\frac{1}{3},\frac{1}{3},-2}$ & 3(3,2)$_{\frac{1}{6},\frac{1}{3},1}$
& 26(1,1)$_{0,0,0}$ &  \\ 
4(\={3},1)$_{\frac{1}{3},-\frac{1}{3},2}$ & 2(\={3},2)$_{-\frac{1}{6},-\frac{1%
}{3},-1}$ &  & \\
\hline
\end{tabular}
\end{center}

As in a usual trinified model, one can see that the fermion content corresponds to the Standard Model (SM) particles plus a right handed neutrino, $N$, plus eleven additional (exotic) particles.

\section{Gauged R-parity and Dark Matter Candidates}

Before discussing R-symmetry, we first dispense with $U(1)_X$ which we break with a VEV for a 
$(1,1)_{0,0,6}$. In \cite{Frampton:2006va} it was shown that if a $Z_2$ R-symmetry is imposed, then one has a natural candidate for dark matter in the lightest conformal particle (LCP).  Here we will argue that this $Z_2$ R-symmetry does not naturally arise in the present model and will not generically be present in any trinified model which originates from orbifolded $AdS_5\times S^5$.  Our reasoning is similar to that found in \cite{martin} where criteria are given for a gauged R-parity surviving in $E_6$ (as well as Pati-Salam and $SO(10)$). The $E_6$ case is similar to ours in that the $E_6$ sub-group examined is $SU_C(3)\times SU_L(2)\times U_Y(1)\times U_{B-L}(1)\times U_X(1)$, and the fermions are in a \textbf{27} of $E_6$ with the same transformations as the fermion content in the model we are examining.  

The current model contains a continuous $U(1)_{B-L}$ symmetry which, when broken by a scalar VEV carrying an even integer amount of 3$(B-L)$, will result in a discrete $(-1)^{3(B-L)}$ symmetry.  This remaining discrete symmetry will then become the $Z_2$ required in \cite{Frampton:2006va} and one will be assured of an LCP dark matter candidate.  One sees that only the scalars $(3,1)_{-1/3,-2/3,-2}$, $(\bar{3},1)_{1/3,2/3,2}$, $(\bar{3},1)_{1/3,2/3,-4}$, and $(3,1)_{-1/3,-2/3,4}$ satisfy the criteria that if one were to obtain a VEV then there would remain the desired discrete R-symmetry. Since any such VEV would break $SU(3)_C$, we conclude that the $Z_2$ R-symmetry can not be arranged in this model, since any other color singlet VEV breaks $B-L$ completely.

\section{Proton Decay}

In typical grand unified trinification models, unification occurs at $M_G \simeq 10^{14}$GeV \cite{willenbrock:2003},
 and rapid proton decay is avoided.    Due to the low scale of unification that can arise in orbifolded AdS/CFT models proton decay must be strictly forbidden by choices of field content and interaction terms.  A mechanism for accomplishing this via baryon charge assignment was put forth in \cite{Frampton:2006pi} where one can assign various baryon numbers to the scalar sector and therefore exclude the unwanted interactions.  We will examine if such a mechanism exists in the present $Z_7$ model.

Once scalar VEVs are obtained, quark and lepton masses are generated from the Yukawa couplings terms 
\begin{eqnarray}
\lambda_q(QH_1\bar{u} + QH_2\bar{d}) + \lambda_l(LH_1\bar{e} + LH_2N)
\end{eqnarray}
where we have defined the scalars $H_1$: $(1,2)_{1/2,0,-3}$ and $H_2$: $(1,2)_{-1/2,0,-3}$.

Additional interactions are seen to exist due to the presence of colored scalars in the model.  In particular there are the Yukawa terms 
\begin{eqnarray}
\lambda_c(QQ\beta + \bar{e}\bar{s}\beta)
\end{eqnarray}
involving the colored scalar that we label $\beta$ which has charges (3,1)$_{-\frac{1}{3},-\frac{2}{3},-2}$.

These are seen to be $B$ and $L$ nonconserving operators and will give rise to the interaction shown in Figure 1.  Once the scalars are integrated out, one will be left with a dimension six effective four-fermion operator.
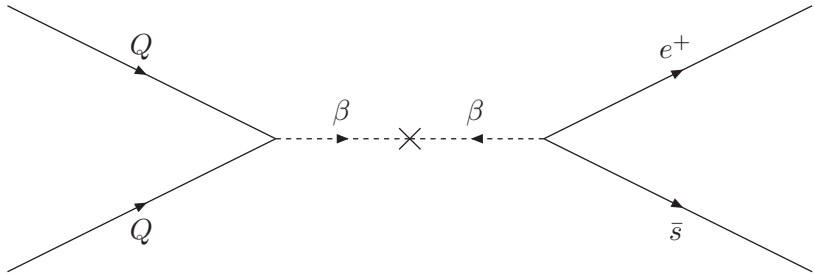
\begin{figure}
\begin{center}
\begin{picture}(300,100)(0,0)
\ArrowLine(0,100)(100,50)
\ArrowLine(0,0)(100,50)
\DashArrowLine(100,50)(150,50){2}
\DashArrowLine(200,50)(150,50){2}
\Line(146,54)(154,46)
\Line(146,46)(154,54)
\ArrowLine(200,50)(300,100)
\ArrowLine(200,50)(300,0)
\Text(50,85)[]{$Q$}
\Text(50,15)[]{$Q$}
\Text(250,85)[]{$e^{+}$}
\Text(250,15)[]{$\bar{s}$}
\Text(125,60)[]{$\beta$}
\Text(175,60)[]{$\beta$}
\end{picture}\\
\end{center}
\caption{Interaction leading to proton decay}
\end{figure}
This is disastrous for the model due to incredibly rapid proton decay via $p \rightarrow K^+ + \nu$.  The rate of proton decay due to this interaction is given by
\begin{eqnarray}
\Gamma = A\frac{1}{m_{\beta}^4}
\end{eqnarray}
where $m_{\beta}$ is the mass of the scalar $\beta$.  The dimensionful proportionality constant, $A$ (of total dimension five), contains relatively well known aspects, given by lattice QCD calculations, as well as model dependent Yukawa couplings.  In a typical trinified model \cite{willenbrock:2006}, proton decay (with reasonable couplings) was suppressed since colored triplet scalars acquire masses on the order of the unification scale, $10^{14}$GeV.  Here one would expect their masses to be on the order of a few TeV and therefore one can immediately conclude that the low scale of unification will produce a proton decay rate in gross disagreement with current experimental bounds.

This has severe consequences for any AdS/CFT orbifold SUSY model with low-scale unification due to the fact that these dangerous colored scalars will always be present in the form of the superpartner of the fermion we have labeled $h$.  This leads us to believe that if a viable conformal model with low scale unification exists, it will not be supersymmetric.

\section{Conclusions}
We have examined the $Z_7$ orbifold of $AdS_5\otimes S^5$ which is the minimal low scale model that can produce the SM fermion content with all three generations present.  It was shown that the scalar content is sufficient to produce the breaking pattern given by $SU(3)^7\rightarrow SU_C(3)\times SU_L(3)\times SU_R(3)\rightarrow SU_C(3)\times SU_L(2)\times U_Y(1)\times U_{B-L}(1)\times U_X(1)$. After breaking $U_{B-L}(1)$ and $ U_X(1)$ it was shown that there exist two Higgs bosons that will provide the usual Yukawa couplings to give the lepton and quark masses and break $SU_L(2)\times U_Y(1)\rightarrow U_{EM}$.  We also shown that there are no scalars in the spectrum to break the continuous $U_{B-L}(1)$ to leave a $Z_2$ R-symmetry which would have provided an LCP dark matter candidate.  This is in contrast to previous work where the $Z_2$ symmetry was implemented by hand.  Proton decay was also examined in the model and found to be disastrous due to the presence of colored scalars which mediates rapid proton decay due to their masses naturally occurring at the TeV unification scale (as opposed to the usual high scale of unification of trinified models which will provide agreement can the current bounds on proton decay).  It was then argued that this is an impasse for any supersymmetric models for this types of orbifolding, since the rapid proton decay will be a feature of any generic low scale model that contains SUSY.  For the non-SUSY case, it still remains possible but challenging to find a model without the problematic scalar fields.    

As the $Z_7$ is the minimal model from which one can obtain a low scale three generation non-supersymmetric realization, evading the proton decay problem due to the existence of colored scalars will be a steep challenge.  As one increases the order of the Abelian orbifold group, one will also increase the particle content and thus the dangerous colored scalars seem to be ubiquitous in models that result in trinification.  One possible avenue to explore would be to examine models of type $SU(N)^n$ with minimal $n$, but with $N$ not equal to 3 and determine if the problem persists.   

 \vspace{-0.4in}

\begin{acknowledgments} 
We thank Paul Frampton for a useful discussion.
  This work was supported by U.S. DoE grant number
  DE-FG05-85ER40226.
\end{acknowledgments}

\end{document}